\documentclass[fleqn,usenatbib]{mnras}

\usepackage{newtxtext,newtxmath}

\usepackage[T1]{fontenc}

\DeclareRobustCommand{\VAN}[3]{#2}
\let\VANthebibliography\thebibliography
\def\thebibliography{\DeclareRobustCommand{\VAN}[3]{##3}\VANthebibliography}

\usepackage{graphicx}	
\usepackage{amsmath}	
\usepackage{float}

\usepackage{textgreek,stackengine}

\newcommand{\Hb}{H$\rm{\beta}$ }

\newcommand{\OII}{[\ion{O}{II}]}
\newcommand{\OIII}{[\ion{O}{III}]}

\usepackage{comment}

\title[Rejuvenation in EoR galaxies]{Rising from the ashes: evidence of old stellar populations and rejuvenation events in the very early Universe}

\author[Witten et al.]{Callum Witten$^{1,2}$\thanks{E-mail: \href{mailto:cw795@cam.ac.uk}{cw795@cam.ac.uk}},
William McClymont$^{2,3}$,
Nicolas Laporte$^{4}$,
Guido Roberts-Borsani$^{5}$, 
Debora Sijacki$^{1,2}$,
\newauthor{Sandro Tacchella$^{2,3}$, 
Charlotte Simmonds$^{2,3}$,
Harley Katz$^{6}$,
Richard S. Ellis$^{7}$,
Joris Witstok$^{2,3}$,}
\newauthor{Roberto Maiolino$^{2,3,7}$,
Xihan Ji$^{2,3}$, 
Billy R. Hayes$^{1}$,
Tobias J. Looser$^{2,3}$,
Francesco D'Eugenio$^{2,3}$
}
\\
$^{1}$Institute of Astronomy, University of Cambridge, Madingley Road, Cambridge CB3 0HA, UK\\
$^{2}$Kavli Institute for Cosmology, University of Cambridge, Madingley Road, Cambridge CB3 0HA, UK\\
$^{3}$Cavendish Laboratory, University of Cambridge, 19 JJ Thomson Avenue, Cambridge CB3 0HE, UK \\
$^{4}$ Aix Marseille Universit\'{e}, CNRS, CNES, LAM (Laboratoire d’Astrophysique de Marseille), UMR 7326, 13388 Marseille, France \\
$^{5}$ Department of Astronomy, University of Geneva, Chemin Pegasi 51, 1290 Versoix, Switzerland \\
$^{6}$ Department of Astronomy \& Astrophysics, University of Chicago, 5640 S Ellis Avenue, Chicago, IL 60637, USA \\
$^{7}$ Department of Physics and Astronomy, University College London, Gower Street, London WC1E 6BT, UK \\
}

\date{MNRAS, submitted}

\pubyear{2024}

\begin{document}
\label{firstpage}
\pagerange{\pageref{firstpage}--\pageref{lastpage}}
\maketitle

\begin{abstract}
While {\it JWST} has observed galaxies assembling as early as $z\sim14$, evidence of galaxies with significant old stellar populations in the Epoch of Reionisation (EoR) -- the descendants of these earliest galaxies -- are few and far between. Bursty star-formation histories (SFHs) have been invoked to explain the detectability of the earliest UV-bright galaxies, but also to interpret galaxies showing Balmer breaks without nebular emission lines. We present the first spectroscopic evidence of a $z\sim7.9$ galaxy, A2744-YD4, which shows a Balmer break {\it and} emission lines, indicating the presence of both a mature and young stellar population. The spectrum of A2744-YD4 shows peculiar emission line ratios suggesting a relatively low ionisation parameter and high gas-phase metallicity. A median stack of galaxies with similar emission line ratios reveals a clear Balmer break in their stacked spectrum. This suggests that a mature stellar population ($\sim 80$~Myr old) has produced a chemically enriched, disrupted interstellar medium. Based on SED-fitting and comparison to simulations, we conclude that the observed young stellar population is in fact the result of a rejuvenation event following a lull in star formation lasting $\sim 20$~Myr, making A2744-YD4 and our stack the first spectroscopic confirmation of galaxies that have rejuvenated following a mini-quenched phase. These rejuvenating galaxies appear to be in an exceptional evolutionary moment where they can be identified. Our analysis shows that a young stellar population of just $\sim 30 \%$ of the total stellar mass would erase the Balmer break. Hence, `outshining' through bursty SFHs of galaxies in the early Universe is likely plaguing attempts to measure their stellar ages and masses accurately.
\end{abstract}

\begin{keywords}
galaxies: high-redshift -- galaxies: ISM -- ISM: lines and bands -- ISM: structure -- cosmology: reionization
\end{keywords}



\section{Introduction}

The advent of the {\it James Webb Space Telescope (JWST)} has facilitated the robust detection of extremely high-redshift galaxies out to $z>10$ \citep[e.g.][]{Curtis-Lake+23,D'Eugenio+24,Finkelstein+23,ArrabalHaro+23,Carniani+24}. The detection of these galaxies pushes back the onset of Cosmic Dawn to at least $z>14$. The identification of galaxies at such redshifts was previously predicted by various studies \citep{Hashimoto+18, Roberts-Borsani+20,Laporte+21,Tacchella+22} due to observed excesses in their optical-continuum detected by {\it Spitzer} relative to the UV-continuum detected with the {\it Hubble Space Telescope (HST)}. This excess was purported to be evidence of an older stellar population and hence an earlier formation redshift that would push Cosmic Dawn to at least $z\gtrsim14$. 

This excess, is primarily associated with the Balmer break, occurring at a rest-frame wavelength of $\sim 3645$\AA. It is strongest at specific temperatures and densities, namely it is most prevalent in A-type stars. However, due to their high temperature, the break is nearly non-existent in O and B stars. In order for a break to be present, the stellar population must have evolved off the main sequence such that A-type stars can dominate the optical continuum. Moreover, nebular emission from ionised gas that is produced thanks to the ionising radiation from extremely young stars can produce an ``inverse Balmer break'' \citep[e.g.][]{Cameron+23b}, which is seen as a decrease in flux between the UV and optical continua, and has been observed to become the dominant feature in high-redshift galaxies \citep{Roberts-Borsani+24}. Thus, the strength of the Balmer break is frequently used as a probe of the timescale of stellar mass assembly and has been well-studied in galaxy formation simulations \citep{Katz+19, Wilkins+24}. 

Therefore, with the detection of $z\sim14$ galaxies by {\it JWST}, one would expect to observe galaxies presenting large Balmer breaks by $z\sim8$, where some of these galaxies could already be $\sim 300$ Myrs old \citep[e.g.][]{Wilkins+24}. Extremely high-redshift, UV-bright galaxies seem to require very bursty star-formation histories \citep[e.g.][]{Carniani+24}, however, the frequency, timescales and rapidity of burst and quenching events is currently poorly understood. 

Recent works have leveraged the significant gains that {\it JWST} has facilitated in imaging the rest-optical wavelengths of high-redshift galaxies to identify robust Balmer break candidates \citep[e.g.][]{Laporte+23,Endsley+23, Trussler+24, Trussler+24b}. Most notably, the use of medium-band filters around the Balmer break wavelength has allowed for the degeneracy between strong nebular emission lines and the Balmer break to be broken \citep{Stiavelli+23, Bradac+24}. This has additionally resulted in vast improvements in stellar mass estimates, finding that most of the Balmer break galaxies sit between $\rm{log}(M_{\star}[M_{\odot}]) = 8-9.5$ \citep{Endsley+23, Trussler+24, Trussler+24b}. 

Some examples of galaxies with potential Balmer breaks, emission lines and extremely red photometry \citep[e.g.][]{Labbe+23} are in a different mass regime ($M_{\star}> 10^{10} M_{\odot}$) than galaxies with bursty SFHs. Moreover, such galaxies have recently been spectroscopically observed by NIRSpec and have been found to host broad-line Active Galactic Nuclei (AGN) \citep{Wang+24}. The contribution of AGN to their spectral energy distribution (SED) makes constraining their SFH challenging. While modelling does suggest the presence of an old stellar population, dynamical mass constraints require that the optical light is dominated by a strong AGN contribution, making the fraction and age of old stars hard to constrain accurately \citep{Wang+24}.

While \textit{spectroscopic} examples of Balmer breaks in galaxies have been seen above $z=7$, these exclusively show no evidence of strong nebular emission lines, indicative of very little recent star formation \citep[e.g.][]{Looser+23,Vikaeus+24, Trussler+24}. These $M_{\star}\sim 10^{8.5} M_{\odot}$ (mini-)quenched galaxies are superb cases for hosting a previous burst in their SFHs, however, they provide only a lower limit on the timescales between bursts. 

While many simulations predicted the bursty SFH of early galaxies \citep{Fukuda+00,Immeli+04, Bournaud+07, Elmegreen+09, Faucher-Giguere+18, Tacchella+20}, observations currently fail to constrain the properties of these bursts, i.e. their duration, frequency and quenching mechanisms. Moreover, direct \textit{spectroscopic} evidence of galaxies rejuvenating on short timescales and hence of mini-quenched periods has yet to be seen. Observational constraints on the burstiness of high-redshift galaxies largely comes from suspected (mini-)quenched galaxies \citep{Looser+23,Dressler+23,Dressler+24,Strait+23,Looser+23b,Trussler+24} and galaxies suspected to be in the process of rapidly quenching \citep{McClymont+24b}, however these are not yet strong enough to constrain models of bursty SFHs. These works suggest that mini-quenched periods lasts at least a few tens of Myrs. This timescale between bursts is of significant interest given that it encodes information about the mechanisms that drive quenching and star-formation events \citep[e.g.][]{Tacchella+20, Dome+23, Kravtsov+24, Sun+23, Mason+23}, which are likely a combination of supernovae and AGN feedback, stellar winds and the effects of mergers and environment. Identifying galaxies that are in a rejuvenating phase, following a previous burst, not only offers the first confirmation that we are observing mini-quenched galaxies in the early Universe, but also offers the perfect laboratory for constraining the burst timescales of early galaxies.

In this context, we aim to utilise spectroscopic observations to identify and analyse rejuvenating galaxies in the early Universe. One of the prime locations for identifying old stellar populations is in the core of protoclusters in the early Universe. Protoclusters undergo a phase of inside-out growth between $z\sim 10-5$, resulting in a stellar mass distribution that is dominant in the central region of the protocluster \citep{Chiang+17}. Existing cores have been seen to host some of the most massive, dust enriched galaxies at early cosmic times \citep{Laporte+17,Laporte+22,Hashimoto+23,Morishita+23}. Just three spectroscopically confirmed protoclusters currently exist at $z>7$: A2744-z7p9OD \citep{Zheng+14,Hashimoto+23,Morishita+23}, SMACS0723-PC \citep[][Witten et al. in prep.]{Laporte+22} and GN-z11-OD \citep{Scholtz&Witten, Tacchella+23}. Only one of these has ALMA dust detections confirming the most massive, dusty galaxy -- A2744-z7p9OD and its constituent A2744-YD4 (hereafter YD4) \citep{Laporte+17}. Therefore, we first introduce the relevant JWST NIRSpec data in Section~\ref{sec:Data}. We then focus on YD4 in Section~\ref{sec:YD4}, which has recently been subject to follow up prism observations as part of the Ultra-deep NIRCam and NIRSpec Observations Before the Epoch of Reionization (UNCOVER) program. With this additional exposure time we have sufficient depth to observe the continuum at rest-frame wavelengths greater than 3600 \AA\ and as such, we are able to analyse the Balmer break of YD4. Using the emission line properties of YD4 we identify a population of similar galaxies and stack their spectra. We discuss the results of this stack in Section~\ref{sec:stack} and use a variety of different models to interpret these results in Section~\ref{sec:Models}. Finally, in Section~\ref{sec:Conclusion} we discuss the implications of these results and our conclusions. 

\section{Data}
\label{sec:Data}
A2744-YD4 was originally identified in HST Frontier Fields imaging \citep{Lotz+17} of the Abell-2744 cluster by \cite{Zheng+14} as part of a $z\sim8$ overdensity, A2744-z7p9OD. Follow up {\it JWST} NIRSpec spectroscopy confirmed YD4 and nine other galaxies in this extreme overdensity \citep{Hashimoto+23,Morishita+23,Cameron+23,Chen+24} at $z=7.87-7.88$. With a comparable overdensity parameter to other high-redshift protoclusters \citep[e.g.][]{Scholtz&Witten,Laporte+22}, A2744-z7p9OD presents as a spectroscopically confirmed protocluster existing in the first 650 millions years of the Universe \citep{Morishita+23}. YD4 resides in the central $11 \times 11$ pkpc core of the protocluster. This region has been recently observed with the {\it JWST} NIRSpec (IFS) leading to the spectroscopic confirmation of 4 galaxies \cite{Hashimoto+23}.
Moreover, Atacama Large Millimeter/ submillimeter Array (ALMA) Band 6 \citep[ID: 2018.1.01332.S, PI: N. Laporte;][]{Laporte+19} and 7 \citep{Laporte+17} observations identify dust continuum in three of these galaxies \citep{Hashimoto+23}, including in the vicinity of YD4 as previously reported by \cite{Laporte+17}. Given the dust continuum detections in the core of the protocluster and the large stellar masses of the constituent galaxies reported by \cite{Hashimoto+23}, the core of this protocluster appears to be evolved. As such the galaxies resident in the core are prime candidates for observing the Balmer break. Previously, \cite{Roberts-Borsani+20} utilised these ALMA detections and {\it Spitzer} observations to identify the presence of an old stellar population within YD4. However, the presence of emission lines reported by \cite{Hashimoto+23, Morishita+23} imply these galaxies are not quiescent and the depth of the original NIRSpec Prism spectroscopy of YD4 and YD7 in the core is insufficient to observe continuum emission at the longer wavelengths where the Balmer break may be found.

The spectroscopic and imaging data exploited, in the case of YD4, were observed as part of the UNCOVER program \citep[ID 2561, PI Labbe;][]{Bezanson+22} and a {\it JWST} Director's Discretionary Time (DDT) program \citep[ID 2756, PI Chen;][]{Roberts-Borsani+23,Morishita+23}. These two programs observed YD4 with the NIRSpec micro-shutter assembly (MSA), using the PRISM/CLEAR configuration. The exposure time of the DDT and UNCOVER observations were 1.2 and 2.3 hours respectively, resulting in a nearly doubling of the signal-to-noise ratio over the DDT data used in \cite{Morishita+23}.

The spectra and NIRCam photometry from each program derive from the reduction and catalogs of \citet{Roberts-Borsani+24}, and we refer the reader to that paper for full details. As a summary, the spectra were reduced using the \texttt{msaexp}\footnote{\url{https://github.com/gbrammer/msaexp}} code, which makes use of a combination of the official STScI JWST pipelines and custom routines for e.g., snowball masking and 1/$f$ removal at the uncalibrated (\textbf{\_uncal.fits}) and count-rate (\textbf{\_rate.fits}) exposure stages. 2D spectra for the object were extracted from the exposures and background subtraction was done using the adjacent MSA shutters forming part of the larger slitlet. The 1D spectrum and associated uncertainties were extracted and combined from the 2D spectra using an optimal extraction procedure, which makes use of a Gaussian kernel fitted to the spatial profile of the spectrum from the (wavelength-)collapsed 2D image. The final 1D spectra from both programs were then combined via an inverse-variance-weighted mean.

The photometric fluxes instead were extracted from 0\arcsec.5 diameter, circular aperture photometry on the reduced and publicly-available images of the \texttt{Grizli} v6 Image Release, based on a NIRCam F115W+F277W+F444W detection image. Using these, we account for slit loss effects that may be present: first we scale the combined NIRSpec spectrum to the rest-frame UV continuum as seen from the NIRCam photometry, and secondly we multiply the spectrum (blueward of the Lyman-break) by a 2nd order polynomial fit to the residual fluxes between the NIRCam and NIRSpec (pseudo-)photometry. Finally, we correct both the spectrum and imaging for lensing magnification, assuming the magnification factor of $\mu=1.96$ listed in Table~3 of \citet{Roberts-Borsani+24}, which use the magnification maps of \cite{Bergamini+23}.

One immediate concern regarding the spectrum of YD4 is that it may be an agglomeration of different sources, especially given its location within an extremely overdense region (see Figure~\ref{fig:spec}). Moreover, in such crowded regions continuum subtraction can become challenging and slit-loss corrections can affect the observed emission line ratios. However, recent JWST NIRSpec integral field spectrograph (IFS) prism observations do not show a radical evolution in the [\ion{O}{III}]$\lambda$5007/\OII\ or [\ion{O}{III}]/\Hb ratios across the spatial extent of YD4 \citep{Venturi+24}, suggesting that the NIRSpec MSA spectrum that we exploit, is unlikely to be an agglomeration of two radically different sources, like has been suggested by \cite{Faisst&Morishita+24}, for the suspected (mini-)quenched galaxy reported by \cite{Looser+23}. Moreover, these emission line ratios are consistent with those that we report for YD4 in Table~\ref{tab:fluxes}, therefore indicating that continuum subtraction and slit-loss corrections to our MSA spectrum appear not to be playing a role in driving the observed emission line ratios. 

The remaining spectroscopic data used in this paper are retrieved in their reduced form from the JWST Advanced Deep Survey (JADES) data release \citep[for galaxies in GOODS-S][]{Bunker+23} and the Dawn JWST Archive (DJA\footnote{\url{https://dawn-cph.github.io/dja/index.html}}; for galaxies in EGS). The DJA data was reduced using the custom-made pipeline MsaExp v.0.6.12 \footnote{\url{https://zenodo.org/record/8319596}}. Further details off the reduction pipeline are given in \cite{Heintz+24}.

\section{A2744-YD4}
\label{sec:YD4}

\begin{figure*}
\centering
\includegraphics[width=1\textwidth]{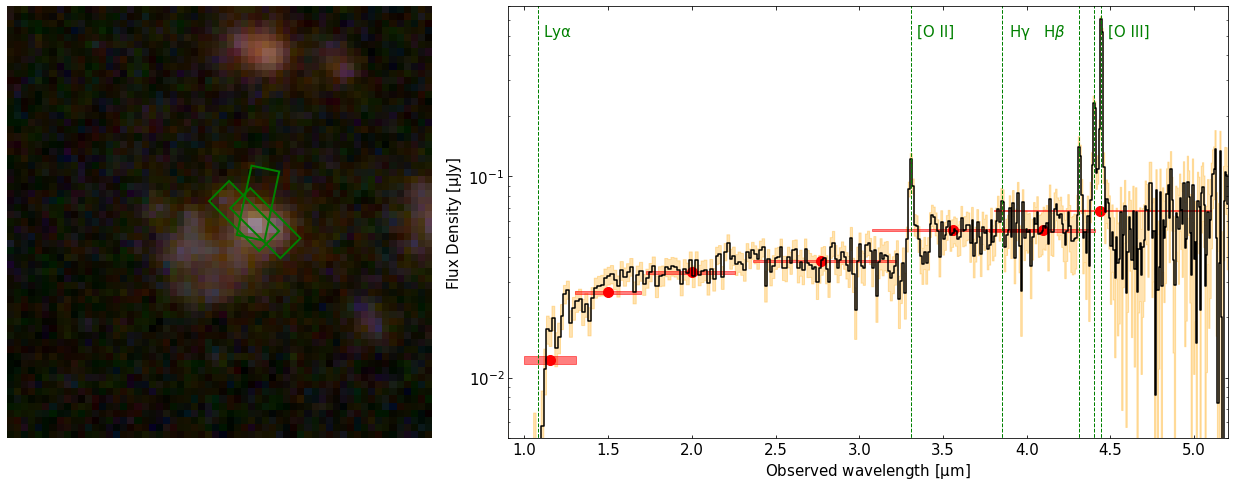}
\caption{(Left) Positions of the three NIRSpec/JWST slit positions used to observe A2744-YD4, the central slit of each observation is indicated in green. For scale, the NIRSpec slit is approximately $0.2^{\prime \prime} \times 0.46^{\prime \prime}$. The NIRSpec observations cover the galaxy YD4 while avoiding significant contamination from YD6 \citep[a seperate $z=7.88$ galaxy to the lower left of YD4;][]{Venturi+24} as discussed further in Section~\ref{sec:Data}. The background image is a color image obtained by combining F150W (blue), F277W(green) and F444W(red), retrieved from the Dawn JWST Archive. (Right) NIRSpec prism spectrum of A2744-YD4 (black) and the 1$\sigma$ uncertainties (yellow). The red points show the NIRCam/JWST photometry and the shaded regions show its uncertainty in the y-axis and the width of the filter in the x-axis. The dashed green lines show the expected positions of notable nebular emission lines. Two key features are seen in the spectrum: a clear Balmer-Break occurring at $\sim 3.3 \mu$m and four emission lines (the blended [\ion{O}{II}]\, doublet, H$\beta$ and the \OIII\, doublet).}
\label{fig:spec}
\end{figure*} 

In Figure~\ref{fig:spec} we present the spectrum of YD4. The spectrum shows two notable features: a Balmer break in the spectrum occurring at $\sim 3.3 \mu$m and four strong nebular emission line features associated with the [\ion{O}{II}]$\lambda$3726,3729~\AA\, (hereafter [\ion{O}{II}]) doublet, H$\beta$ and both components of the [\ion{O}{III}]$\lambda$4959,5008~\AA\, (hereafter[\ion{O}{III}]) doublet. 

\begin{table}
    \centering
    \caption{}
    \begin{tabular}{lc}
        \hline
    Parameter & Measurement $^{*}$ \\
    \hline
    Fluxes [$10^{-19}$ ergs/s/$\rm{cm}^{2}$] & \\
    $\rm{[\ion{O}{II}]}$ & $7.3 \pm 0.1$\\
    \Hb & $2.3 \pm 0.2$\\
    $\rm{[\ion{O}{III}]\,}_{4959}$ & $5.8 \pm 1.1$\\
    $\rm{[\ion{O}{III}]\,}_{5007}$ & $15.1 \pm 0.2$\\
    &\\
    EW$_{0}$ [\AA] & \\
    \Hb & $25 \pm 3$\\
    $\rm{[\ion{O}{III}]\,}_{5007}$ & $170 \pm 18$\\
    &\\
    Line ratios & \\
    log$_{10}$(O32) & $0.12 \pm 0.01 $\\
    log$_{10}$(R23) & $1.18 \pm 0.03 $\\
    \hline
    \end{tabular}
    \\
    Notes:$^{*}$ All measurements have been de-magnified assuming $\mu = 1.96$ from \cite{Bergamini+23}. The line ratios are dust corrected using the dust attenuation law obtained from our \texttt{PROSPECTOR} modelling in Section~\ref{sec:SED-fitting}. 
    \label{tab:fluxes}
\end{table}

\begin{figure}
\centering
\includegraphics[width=0.5\textwidth]{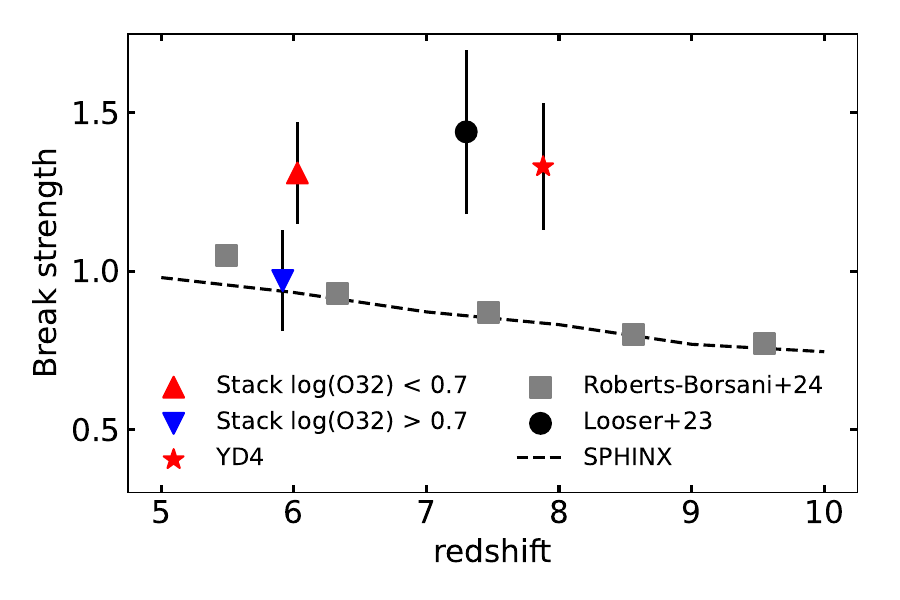}
\caption{The Balmer break strength, $F_{\nu, 4225}/F_{\nu, 3565}$, using the definition of \citet{Roberts-Borsani+24}, showing the break strength in YD4 (star) and our two stacks (triangles). These are compared to the median break strength found in the SPHINX$^{20}$ simulations \citep[dashed line;][]{Rosdahl+18,Rosdahl+22, Katz+23} and stacked NIRSpec Prism spectra from \citet{Roberts-Borsani+24} (squares). We also show the break strength of the galaxy from \citet{Looser+23} (circle).}
\label{fig:Balmer_break_strength}
\end{figure}

\begin{figure*}
\centering
\includegraphics[width=1.0\textwidth]{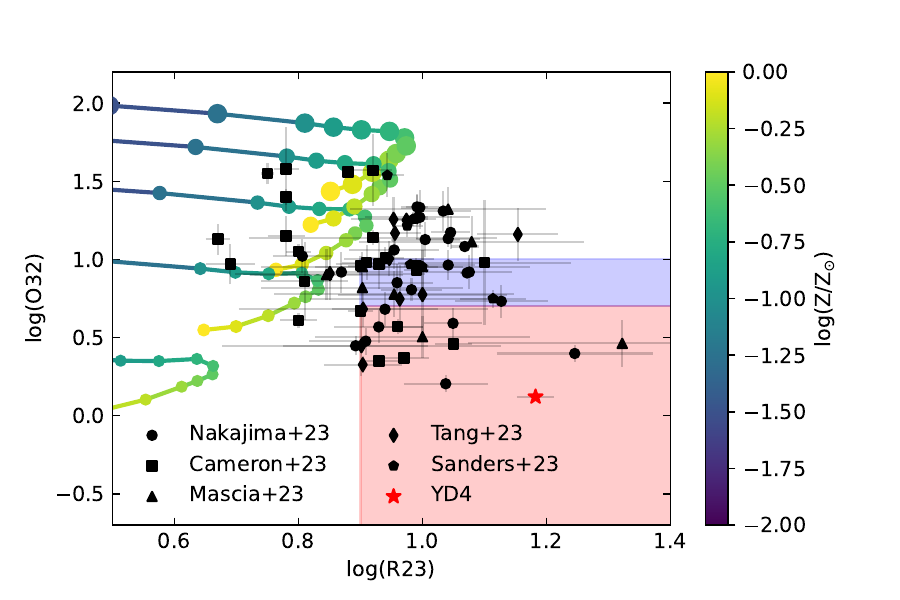}
\caption{The dust-corrected excitation-ionisation plot, showing NIRSpec data from \citet{Nakajima+23, Cameron+23, Mascia+23, Tang+23, Sanders+23}. We additionally show the evolution of the O32 and R23 ratios with metallicity (colourbar) and ionisation parameter (increasing marker size, from -3 to -1, in steps of 0.5) using our 5 Myr \texttt{CLOUDY} models \citep{Ferland+13}, discussed further in Section~\ref{sec:CLOUDY}. The values for YD4 are denoted by a red star, while the two stack criteria are indicated by blue and red boxes. Clearly YD4 resides in a poorly populated region of the excitation-ionisation plane.}
\label{fig:exc_vs_ion}
\end{figure*}

While observations of Balmer breaks in galaxies at $z>7$ exist \citep[e.g.][]{Looser+23}, they are certainly rare. \cite{Roberts-Borsani+24} have shown instead that inverse Balmer breaks, so-called Balmer jumps, appear to be typical in galaxies above $z\sim6$, as shown in Figure~\ref{fig:Balmer_break_strength}. We choose to exploit the Balmer break definition of \cite{Roberts-Borsani+24}, which is more restrictive than typical break definitions \citep[e.g.][]{Binggeli+19}. This definition, $B=F_{\nu}(4225 $\AA$)/F_{\nu}(3565 $\AA$)$, uses windows covering $3500 - 3630$ \AA\, and $4160 - 4290$ \AA\ thus avoiding any potential strong emission lines, such as the [\ion{Ne}{III}]$\lambda$3869,3967~\AA\,  doublet and \ion{He}{I} ($\lambda$3889~\AA) emission lines. This results in a measured break strength of $B=1.3\pm0.2$ for YD4, where $B>1$ is a traditional Balmer break, while $B<1$ indicates significant nebular emission producing an inverse Balmer break. When we compare this break strength to those presented in \cite{Roberts-Borsani+24}, in Figure~\ref{fig:Balmer_break_strength}, it becomes clear that the break observed in YD4 is exceptional relative to galaxies at a similar redshift and is comparable to that of the suspected (mini)-quenched galaxy from \cite{Looser+23} ($B=1.4\pm0.3$). 

As we discussed earlier, UV-optical breaks have been observed in AGN-host galaxies. These breaks can be a combination of emission from the accretion disk of the AGN and also a Balmer break \citep{Wang+24}. As such, given that protocluster cores appear ideal locations to form and fuel massive black holes \citep{Bennett+24} and that YD4 appears to be evolved thanks to its ALMA dust detections \citep{Laporte+17}, one could postulate that the break observed in YD4 may be driven by an AGN. However, it is crucial to note that, in galaxies that have optical continua seemingly dominated by AGN accretion disk emission (e.g. the so-called little red dots), the slope of their optical continua is much more red \citep[e.g.][]{Labbe+23b,Furtak+24, Wang+24, Matthee+24,Kocevski+24} than the flat optical slope seen in Figure~\ref{fig:spec}. Therefore, given YD4 does not exhibit a steep, red optical continuum, evidence of emission lines that require AGN photo-ionisation \citep[e.g.][]{Scholtz+23} or of broad Balmer emission lines, we conclude that there appears to be no evidence that YD4 is in an AGN phase.

We fit the four emission lines seen in the spectrum of YD4 with Gaussian profiles. This fitting occurs in $f_\nu$ space where the optical continuum is flat, we therefore estimate the continuum level of the Gaussian by averaging over the local optical continuum while masking strong nebular emission lines. The resulting flux of these lines is reported in Table~\ref{tab:fluxes}. 

Figure~\ref{fig:exc_vs_ion} shows the dust-corrected emission lines ratios from multiple JWST programs: JADES \citep{Bunker+23}, the Cosmic Evolution Early Release Science \citep[CEERS;][]{ArrabalHaro+23}, Early Release Observations (ERO) and GLASS \citep{Treu+22}. In order to compare our observed ratios to these, we dust-correct the emission line fluxes reported in Table~\ref{tab:fluxes} using the dust properties obtained by \texttt{PROSPECTOR} in Section~\ref{sec:SED-fitting}. One of the particularly noteworthy features about the emission lines seen in YD4 is the relatively low dust-corrected $\OIII \lambda5007/\OII$ ratio (hereafter O32) of $\rm{log}_{10}(\rm{O32}) = 0.12 \pm 0.01$. In combination with the high dust-corrected (\OIII + \OII)/\Hb ratio (hereafter R23) of $\rm{log}_{10}(\rm{R23}) = 1.18 \pm 0.03$, these ratios indicate that YD4 resides in a poorly populated region of the excitation-ionisation plane, shown in Figure~\ref{fig:exc_vs_ion}. 

The R23 ratio is often taken as a tracer of the gas-phase metallicity \citep{Curti+23}, while O32 is a well-known tracer of ionisation parameter \citep{Kewley+19}. Therefore, these ratios seen from YD4 are indicative of a relatively low ionisation parameter yet a relatively high metallicity. The low O32 ratio appears consistent with a stack of $z\sim2-3$, $\rm{log_{10}(M_{\star}[M_{\odot}])}\sim 9.5$ galaxies from the MOSDEF survey in \cite{Sanders+21}. This suggests that YD4 may be more evolved than the extreme O32 ratio galaxies that have been seen in the early Universe. These results therefore pose the question of whether the presence of a Balmer break is indeed connected to observing a low O32 ratio and a high R23 ratio. 

\section{Stack}
\label{sec:stack}
In the following section we analyse galaxies that lie at a similar position on the R23-O32 plane, to establish whether they also host old stellar populations. However, continuum detections at $\lambda_{\rm{rest}}>0.36\mu$m remain challenging, even in the era of JWST. The decrease in the sensitivity of NIRSpec with increasing wavelength often makes identifying Balmer breaks in galaxies difficult. While deeper observations with future JWST programs will facilitate the confirmation of Balmer breaks in candidate galaxies, however, these observations are not currently available. We therefore employ a stacking technique, allowing us to identify the presence of a Balmer break within a carefully selected sample of galaxies with similar emission line properties to YD4.

We utilise all galaxies in the public NIRSpec Prism observations by the JADES and CEERS surveys of the GOODS-South and EGS fields. We select our sample based on dust-corrected emission line fluxes taken from \cite{Cameron+23,Tang+23,Nakajima+23,Mascia+23, Sanders+23}. We select all galaxies at $z>5.5$, observed with NIRSpec prism, located in the bottom right region of the excitation-ionisation plot, shown in Figure~\ref{fig:exc_vs_ion}, similar to YD4:
\begin{itemize}
    \item $\rm{log}_{10}\rm{O32} < 0.7$
    \item $\rm{log}_{10}\rm{R23} > 0.9$\,.
\end{itemize}

We identify a total of 10 galaxies between $5.5 < z < 8.7$ that satisfy these conditions: JADES-GS+53.17582–27.77446, JADES-GS+53.15613–27.77584, JADES-GS+53.11911–27.76080, JADES-GS+53.12259–27.76057, JADES-GS+53.13002–27.77839 \citep{Cameron+23}, CEERS-1029, CEERS-1023, CEERS-1102 \citep{Tang+23}, CEERS-00403, CEERS-00545 \citep{Nakajima+23}.

We first shift each galaxy's spectrum to a common rest-frame wavelength grid, with a resolution of 0.001~$\mu$m by exploiting the SpectRes code from \cite{Carnall+17}. We then normalise all spectra by their UV flux at $0.15 \mu$m and perform a median stacking. The efficacy of the median stacking approach has been shown by numerous previous works \citep[e.g.][]{Witten+23, Roberts-Borsani+24, Harikane+20}. We estimate the uncertainty in our spectra using bootstrap sampling. We produce 1,000 median stacks by randomly selecting $N$ galaxies from our sample allowing for replacement, where $N$ is the number of galaxies in each sample, and median stacking the spectra of the selected galaxies. When a galaxy is selected to be included in the stack we redraw the galaxy's spectrum from Gaussian's centred on the flux in a given wavelength bin and the standard deviation of the Gaussian is given by the uncertainty in the flux. We calculate the median and standard deviation of the 1,000 stacked spectra and report this as the final median stacked spectrum and its associated uncertainty. 

\begin{figure}
\centering
\includegraphics[width=0.5\textwidth]{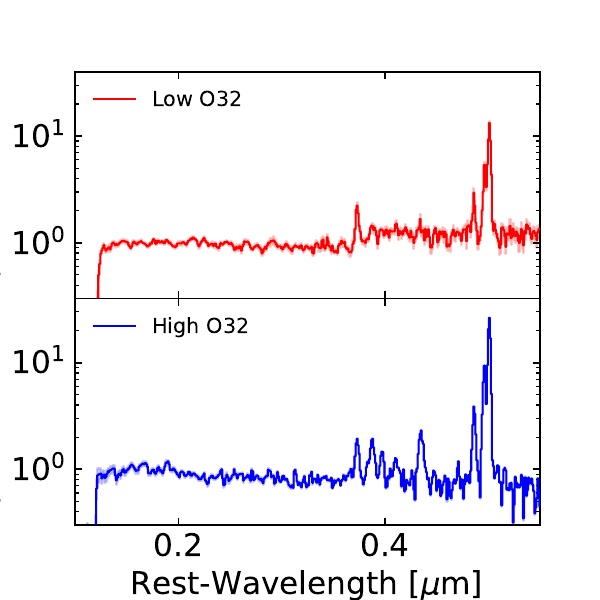}
\caption{The median stacked spectra of our two selection criteria, shown in Figure~\ref{fig:exc_vs_ion}, where red (top panel) indicates those selected with a lower O32 ratio and blue (bottom panel) are those with a higher O32 ratio. The higher ionisation parameter stack shows the presence of multiple optical emission lines and indications of multiple UV emission lines, unlike the lower ionisation parameter stack. The low ionisation parameter stack shows a strong Balmer break, contrary to the flat UV-optical continuum seen in the high ionisation parameter stack.}
\label{fig:high_vs_low_logU}
\end{figure}

The final stacked spectrum can be seen in Figure~\ref{fig:high_vs_low_logU}. A Balmer break can clearly be identified and we find a break strength of $B=1.3\pm0.2$. This result shows that a low O32 ratio is linked to the presence of a Balmer break. The association between a Balmer break and low O32 ratio is curious because they are each associated with different stellar ages. The Balmer break is produced by older stars ($\gtrsim$50\,Myr) whereas strong nebular emission is driven by young stars ($\lesssim$10\,Myr). This may imply some direct causal relationship between a previous generation of stars and the young stars, such as metal enrichment of the ISM, or that the older stars are a good proxy for another galactic property. We investigate this association further in Section~\ref{sec:Models}.

It is important to note that we are also selecting based on the R23 ratio. Therefore, in order to confirm that the discovery of a Balmer break is not only driven by a metallicity dependence, and hence largely driven by R23, we also stack galaxies with our R23 selection, but with $0.7 < \rm{log_{10}(O32)} < 1.0$, contrasting our typical cut of $\rm{log_{10}(O32)} < 0.7$. The selection criteria can be seen in blue in Figure~\ref{fig:exc_vs_ion} and correspond to a higher ionisation parameter, while falling below the most extreme O32 ratio galaxies seen by JWST. Following the same stacking procedure as discussed above in Section~\ref{sec:stack}, we find very different spectral properties. The higher O32 and hence higher ionisation parameter stack reveals a multitude of rest-frame UV and optical emission lines as well as a UV-optical continuum that is consistent with no break ($B=1.0\pm0.2$), as shown in Figure~\ref{fig:high_vs_low_logU}. The strength of the Balmer break is shown in Figure~\ref{fig:Balmer_break_strength} and appears consistent with a stack of star-forming galaxies at similar redshifts from \cite{Roberts-Borsani+24} and the median galaxy from the SPHINX$^{20}$ simulation (discussed further in Section~\ref{sec:SPHINX}). This strongly implies that the presence of a Balmer break in the high O32 sample is linked to both relatively high metallicity and low ionisation parameter. 

\section{Modelling}
\label{sec:Models}
In the previous section we have shown that the emission line ratios and Balmer break in YD4 are exceptional when we compare to other high-redshift galaxies. Moreover, when we select for these peculiar emission line ratios in our stack sample, we discover a Balmer break in the median stack. We have justified that, based on its emission line properties, YD4 appears not to be in an AGN phase and that the observed break cannot be explained by accretion disk emission, and for the same reasons, this is also true for our stacked spectrum. Therefore, in the following section we exploit multiple forms of stellar population modelling to constrain the relationship between the SFH and the observed emission line ratios.

\subsection{Simple stellar population analysis}
\label{sec:CLOUDY}
While we identify a Balmer break in our sample, it remains unclear how the break is connected to the production of the observed O32 and R23 emission line ratios. In order to understand the properties of the stellar populations that are able to produce these ratios we employ the photoionisation code \texttt{CLOUDY}. We create grids using the stellar evolution models BPASSv2.2.1 \citep{Eldrige+17, Stanway+18} including binaries and a \cite{Chabrier+03} initial mass function with a high-mass cut off of $300 \rm{M}_{\odot}$. We use stopping conditions based on electron fraction $e_{\rm frac}=0.01$ and temperature $T=3,000$K. We assume spherical geometry with an inner radius of $0.1$~pc. We vary the ionisation parameter, metallicity, stellar age and density over the given ranges:
\begin{itemize}
    \item log$U$ = [-3.0, -2.5, -2.0, -1.5, -1.0]
    \item log$_{10}$(Z/Z$_{\odot}$) = [-2 : -1] in steps of 0.25, 
    [-1 : 0] in steps of 0.1
    \item ages [Myr] = [1, 3, 5, 7, 10, 25, 50, 100]
    \item log$_{10}$($\rho [\rm{cm}^{-3}]$) = [2 , 3, 4]\,.
\end{itemize}

We note that the failure of photoionisation models to recreate the O32 and R23 emission line ratios at a sufficient strength as to match observations has often been noted at high redshifts \citep[e.g.][]{Roberts-Borsani+24, Cameron+23}. One potential solution to this is to account for turbulence when modelling emission line ratios. \cite{Gray&Scannapieco} find that when they include a turbulent velocity of $\sim 40$ km/s in their modelling they can increase the R23 ratio by $\sim 0.3$ dex which would in turn shift our \texttt{CLOUDY} models into the shaded region in Figure~\ref{fig:exc_vs_ion}. While our simple \texttt{CLOUDY} models also fail to reproduce the observed O32 and R23 ratios for YD4 and our sample of galaxies at any of the considered stellar ages and densities, extrapolating from trends shown in Figure~\ref{fig:exc_vs_ion} allows us to understand the likely required properties of YD4 and our stack. As discussed in Section~\ref{sec:YD4}, O32 is a well known tracer of ionisation parameter, and the low O32 observed in YD4 and our stack galaxies relative to the general population implies that the galaxies we study in this paper have a low ionisation parameter of around log$U$$\sim -2.5$. The \texttt{CLOUDY} tracks show a turnover with increasing metallicity, occurring at a lower R23 ratio than that seen in our sample of galaxies, however, given the overall trend of increasing metallicity with R23 ratio and a higher R23 ratio in our galaxies than the general population, we conclude that our sample appears to be more chemically enriched than the general population and that the metallicity is likely log$_{10}(Z/Z_{\odot})\sim -0.5$ -- the values observed at the turnover of R23. We also note that this enrichment is reflected in the observed ALMA dust detection for YD4 and the high dust attenuation predicted by \texttt{PROSPECTOR}, in Section~\ref{sec:SED-fitting}, for both YD4 and the stack. This is similar to the conclusion of \cite{Cameron+23} who suggest that their galaxies with the highest R23 and lowest O32 appear to be consistent with the turnover in the excitation-ionisation plane seen at $z\sim0$ \citep{Curti+23}. \cite{Cameron+23} conclude that this suggests that galaxies with these emission line ratios are likely the most chemically enriched in their sample, with metallicities of $0.3-0.5 Z_{\odot}$, in agreement with our conclusion.

Figure~\ref{fig:exc_vs_ion} shows the tracks for \texttt{CLOUDY} models with a stellar age of 5 Myrs and a density of 10$^{4} \rm{cm}^{-3}$. However we note that while changing these parameters can make differences in the exact position of the tracks, often along the direction of changing ionisation parameter, they do not help to reproduce the observed ratios and do not significantly affect the implied metallicity and the relatively low ionisation parameter. The main change in the emission lines is to their equivalent width (EW) which significantly decreases with age. 

Previous modelling has shown that the strength of the [\ion{O}{III}]\, emission line, the strongest optical emission line seen in the spectra of YD4 and our stack, falls rapidly with the increasing age of a stellar population. \cite{Faisst&Morishita+24} have shown that the [\ion{O}{III}]\, emission line luminosity produced by a stellar population decreases by more than two orders of magnitude after just 5 Myrs due to the population's decreasing ionising emissivity. The detection and strength of the optical emission lines in YD4 and the stack therefore imply that these galaxies host stellar populations with ages of $\lesssim 10$~Myrs.

\begin{figure}
\centering
\includegraphics[width=0.5\textwidth]{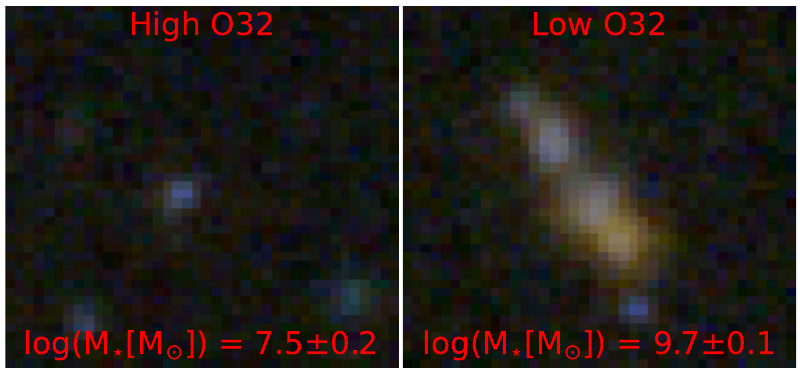}
\caption{Colour images of two galaxies selected from each of the high O32 (left) and low O32 (right) samples from Section~\ref{sec:stack}. The images were obtained by combining F150W (blue), F277W(green) and F444W(red), retrieved from the Dawn JWST Archive. The stellar masses reported are taken from \citet{Morishita+24}. The high O32 galaxy is more compact and has a lower stellar mass than the low O32 sample, as well as exhibiting bluer colours.}
\label{fig:Images}
\end{figure}

While the production of the emission lines by a young stellar population is not surprising, the requirement for a relatively low log$U$ and relatively high metallicity, compared to other high-redshift galaxies \citep[e.g.][]{Cameron+23}, is noteworthy. Significant previous work has been undertaken in understanding the trend of increasing ionisation parameter with redshift \citep[e.g.][]{Brinchmann+08, Bian+16, Papovich+22,Liu+08, Masters+16}. This previous work targeting the evolution in log$U$ by $z\sim2-3$ indicates that $U \propto \Sigma_{\rm{SFR}}$, the star formation rate surface density, and this relation has been linked, at least in part, to increasing electron density \citep{Reddy+23a,Reddy+23b} which is known to evolve with redshift \citep[e.g.][]{Isobe+23}. Unfortunately, the resolution of NIRSpec prism observations makes resolving the \OII\ doublet, a key diagnostic of electron density \citep[e.g.][]{Sanders+16}, impossible. However, thanks to analysis by \cite{Morishita+24}, we can evaluate the $\Sigma_{\rm{SFR}}$ using their reported SFRs and effective radii. We find no reduction in the $\Sigma_{\rm{SFR}}$ of our low O32 sample relative to the higher O32 sample in Section~\ref{sec:stack}. However, this relationship has large scatter \citep{Reddy+23a,Reddy+23b} and our results are likely driven by the minimal overlap between our samples and those of \cite{Morishita+24}, thus producing a very small sample size. Moreover, our higher O32 sample does not include the most extreme O32 ratios seen in Figure~\ref{fig:exc_vs_ion}. 

The galaxies in our sample that do overlap with \cite{Morishita+24} appear to have larger effective radii and stellar masses than both the higher O32 sample, and the general galaxy population at their given redshift. We see a qualitative trend towards more extended structures from NIRCam imaging of all of our low O32 sample. They also show much redder colours than the high O32 stack, relating to their much redder UV slope caused by their more evolved state, an example of which can be seen in Figure~\ref{fig:high_vs_low_logU}.

We therefore interpret the reduced O32 ratio in our sample as evidence of an enriched ISM and a low ionisation parameter. This is likely driven by a lower SFR surface density, and therefore lower electron density \cite{Reddy+23b}. Given the association we find between the low O32 ratio and the presence of a Balmer break, these ISM conditions seem to be driven by the presence of an old stellar population that has chemically enriched the ISM, via supernovae events. Additionally, the Balmer break indicates the presence of an older stellar population and hence a more evolved galaxy which may be more extended and transitioning away from the bursty mode of star formation, and thus creating a reduced SFR surface density \citep[see e.g.][]{Hopkins+23, McClymont+24b}.

However, in order for a full analysis of the properties driving the low O32 ratio, a larger sample, and high resolution spectroscopy of the \OII\ doublet are required. As it stands, our sample is currently too small to make any such conclusions, however, future JWST programs aimed specifically at observing candidate (mini-)quenched galaxies, but moreover, general large surveys, will inevitably unveil a larger population of rejuvenating galaxies.

\subsection{SED fitting}
\label{sec:SED-fitting}

The detection of the Balmer break within the spectra of YD4 and our stack is noteworthy, not just because it indicates that an old stellar population is present, but implies that the young stellar population is not dominating the SED of these galaxies, likely indicating the presence of very few, very young stars. Constraining what star-formation history is capable of producing these features, and understanding the final chemical and dust enrichment is crucial in understanding the link between the O32 and R23 line ratios and the Balmer break seen in YD4 and our stack. 

In this context, we expand our analysis of YD4 and our stacked spectrum by fitting them with the SED-fitting code \texttt{PROSPECTOR} \citep{Prospector}. We chose to exploit \texttt{PROSPECTOR} as a modelling tool thanks to its flexible dust attenuation curve and use of BPASS models. The use of BPASS models \citep{Eldridge&Stanway} are especially noteworthy in the context of the ionising output of young stellar populations, where the ionising photon emissivity declines rapidly for the \cite{Bruzual&Charlot} single-star models while this decline is more gradual for BPASS models, when we account for the impact of binary stars. As such, binary evolution can have a strong impact on the nebular emission modelling \citep{Xiao+18, McClymont+24} and is therefore important for constraining the exact SFH of galaxies that have ongoing star formation, like those studied in this paper. We fit both the photometric data and the spectroscopic constraints on the emission line fluxes (seen in Table~\ref{tab:fluxes} for YD4) following a similar approach to \cite{Tacchella+23}. We utilise a non-parametric SFH from \citet{Leja+19}. The photometric constraints of YD4 were obtained in the framework of the UNCOVER survey (GO 2561, P.I.:Ivo Labbe) and PSF matched to F444W \citep{Weaver+24}, while for the stack we artificially produce the photometry by passing the spectrum through the transmission curves of the JWST filters. We fix the redshift of both YD4 and the stack at the spectroscopic redshift measured in \citet{Morishita+23} ($z=7.88$) and re-normalise the stack to have the same UV luminosity as YD4. This is motivated by our intention to place the SFHs on comparable scales, we note this will result in stellar masses and formation redshifts that are not representative of the median galaxy in our stack sample, but allows us to make comparisons between the shapes of the SFHs. 

We make use of a similar model to that described in \cite{Tacchella+23}, employing a 16-parameter physical model and priors. We fit for the stellar mass, stellar metallicity, gas-phase metallicity, ionisation parameter and dust attenuation. Our dust attenuation modelling includes three-components -- a diffuse component over the entire galaxy, an additional component attenuating the light from stars less than 10 Myrs old and finally a flexible slope, following the prescription of \cite{Conroy+09}. Our non-parametric SFH employs a continuity prior, described by a Student's t-distribution with $\sigma = 0.3$ and $\nu = 2$, on the relative star formation in each adjacent age bin. We fit nine parameters that control the ratio of the average star-formation rate within ten adjacent bins. These bins span the lookback time range of $0-284$ Myrs. We chose to use a coarse binning in time to avoid an over-interpretation of our results with \texttt{PROSPECTOR}. While the emission lines allow constraints on the star formation on $\sim 10$ Myr timescales, the Balmer break constrains star formation on larger timescales ($\sim 100$ Myr), meaning information is only available to constrain the SFH on such scales. As such we exploit bins first covering 0-5 Myrs and 5-10 Myrs time range, while the remaining eight bins are log-spaced to $z=12$ (284 Myrs). Therefore, when we make statements about the duration of burst and quenched periods, these should be taken as approximate values, noting the bin widths covering these periods. 

\texttt{PROSPECTOR} models nebular emission line fluxes using the Flexible Stellar Population Synthesis code \citep[FSPS; see][]{Byler+17}, which in turn is based on \texttt{CLOUDY} \citep{Ferland+13} photoionsation models. This is notable as we show in Figure~\ref{fig:exc_vs_ion} that \texttt{CLOUDY} modelling is unable to reproduce the dust-corrected emission line ratios with a single stellar population. However, the more complex agglomeration of stellar populations that \texttt{PROSPECTOR} employs is able to reproduce the observed emission line fluxes within their associated uncertainties.

The best fit model of YD4 is obtained with  the following set of parameters: stellar mass $M_{\star}$=(4.1$^{+0.2}_{-0.2}$)$\times10^{9}\ \rm{M_{\odot}}$, stellar metallicity log$_{10}Z_{\star}/Z_{\odot} = -2.84 ^{+0.1}_{-0.09}$, gas-phase metallicity log$_{10}Z_{\rm{gas}}/Z_{\odot} = -0.31 ^{+0.01}_{-0.01}$, ionisation parameter log$U$ $= -2.49 ^{+0.01}_{-0.01}$, diffuse dust V-band optical depth $\tau = 0.67^{+0.06}_{-0.03}$. The inferred metallicity and ionisation parameter values are in agreement with the values that we predicted from our \texttt{CLOUDY} modelling. 

We do not report all of the best fit parameters derived for the fitting of the  stack, as our choice of redshift and normalisation, to match the spectrum of YD4 means properties like the stellar mass do not relate to the actual stellar mass of the median galaxy in our stack. However, we note that the implied ionisation parameter (log$U$ $= -1.93 ^{+0.07}_{-0.05}$), metallicity (log$_{10} Z_{\rm{gas}}/Z_{\odot} = -0.52 ^{+0.01}_{-0.01}$) and dust ($\tau = 0.96^{+0.02}_{-0.04}$) are consistent with the values seen in YD4. These results again strongly imply that these specific values for these parameters are intricately linked to the position in the excitation-ionisation plot and moreover are affected by the presence of an old stellar population, given the inferred SFH discussed below. 

\begin{figure}
\centering
\includegraphics[width=0.5\textwidth]{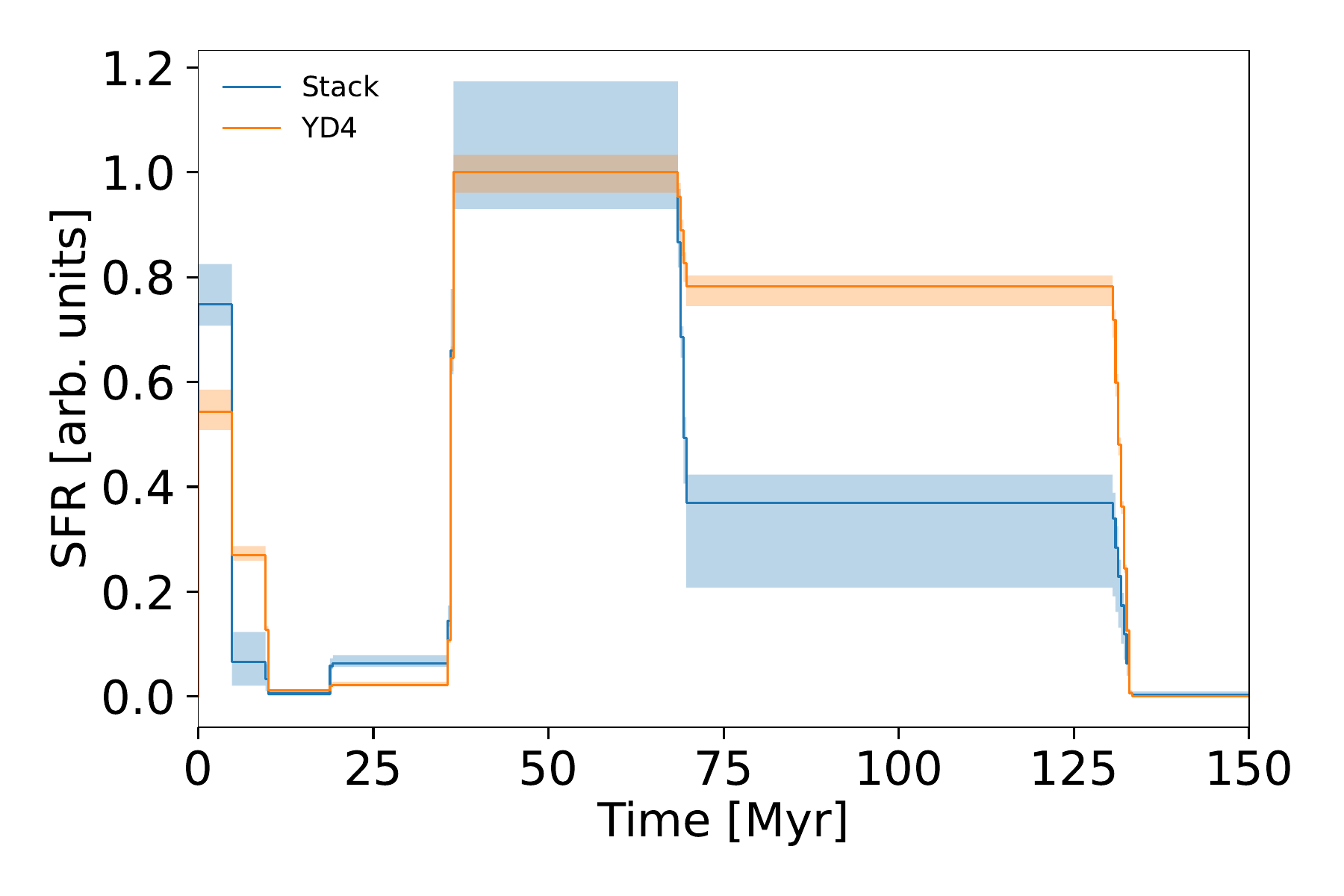}
\caption{The SFH derived from our \texttt{PROSPECTOR} fitting of the photometry and emission lines of YD4 (orange) and our stacked spectrum (blue). The average SFR within each SFH bin is shown as a flat line across the width of the age bin. The star-formation rates have been normalised by the maximum SFR in the SFH for ease of comparison.}
\label{fig:SFHs}
\end{figure}

The best fit SFH and its uncertainty, achieved by our \texttt{PROSPECTOR} fitting  of both YD4 and our stacked spectrum, can be seen in Figure~\ref{fig:SFHs}. These SFHs have been rescaled so that their maximum SFRs are unity. This allows us to compare the shapes of their SFHs. In the SFH of YD4, the first star-forming event seen is a major burst beginning $\sim 125$ Myrs pre-observation and lasting for nearly 100 Myrs. We then see a 25 Myr period of quiescence, followed by a burst of star formation that reaches $60\%$ of the maximum star formation rate, but its relatively short duration means it forms just $5\%$ of the total stellar mass. 

We remind the reader that our investigation of the SFH of the stack is motivated as a comparison to YD4. The shape of the SFH of the stack is notably similar to YD4, with a significant burst finishing $\sim 35$ Myrs in the past, followed by a lull in star formation. Finally, a recent ($t < 5$ Myrs) burst of star formation is seen. This recent burst is at $75 \%$ of the maximum SFR seen in the stack and has a slightly smaller old-stellar population than YD4 forming fewer stars $>75$ Myrs ago. However, the recent burst in the SFH of the stack still only forms $6\%$ of the total stellar mass.

\begin{figure}
\centering
\includegraphics[width=0.48\textwidth]{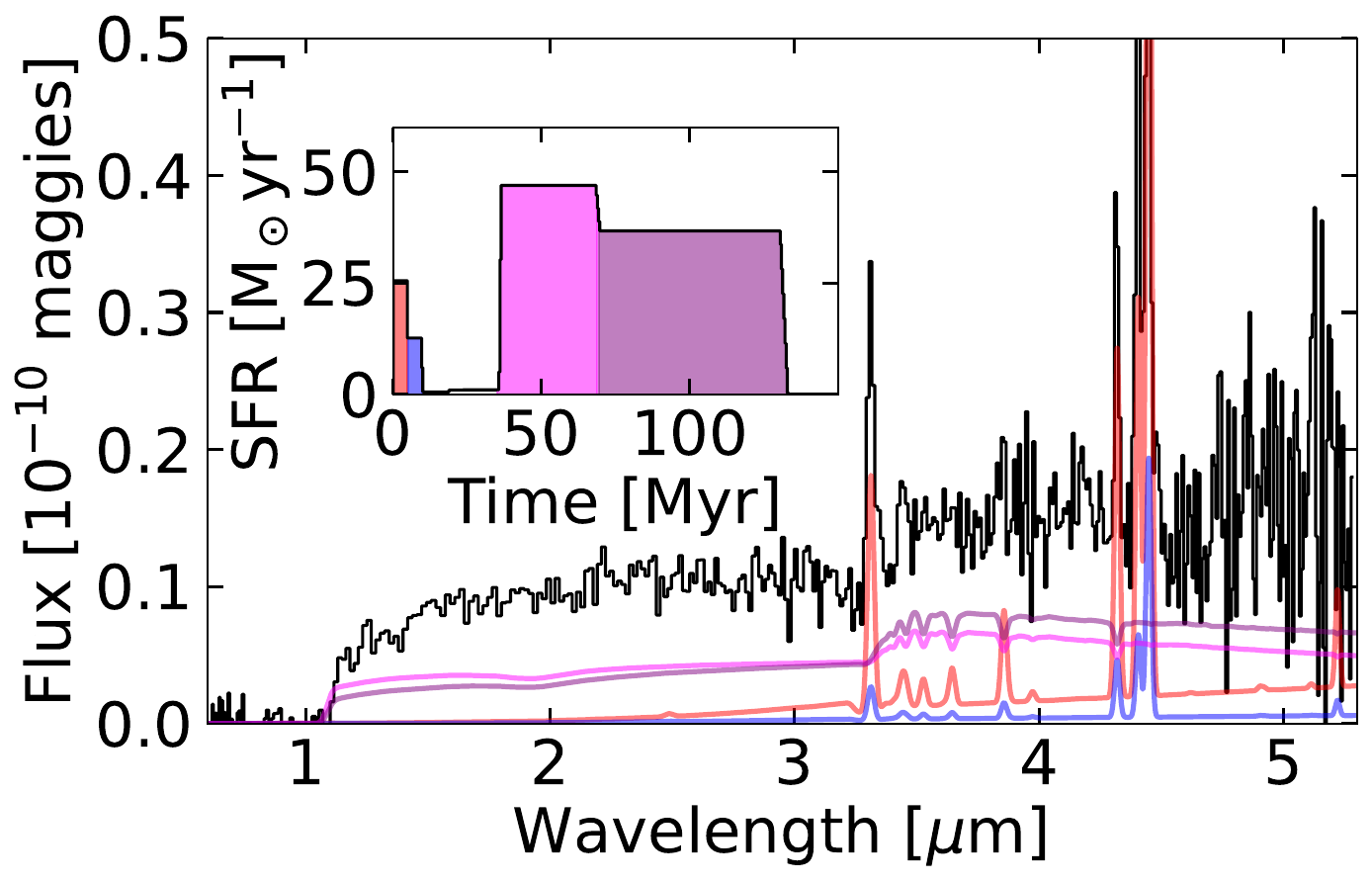}
\caption{The NIRSpec Prism spectrum of YD4 shown in black. The SFH inferred from \texttt{PROSPECTOR} is shown in the inset panel. The components of the spectrum attributed to the main star formation bins are shown by different colours: 0-5 Myrs (red), 5-10 Myrs (blue), 35-70 Myrs (pink) and $>75$ Myrs (purple). The old stellar population ($>35$ Myrs) dominates the continuum emission, while the young population (0-10 Myrs) produces all of the observed nebular emission lines.}
\label{fig:StellarPops}
\end{figure}

In Figure~\ref{fig:StellarPops} we consider the relative contributions to the overall spectrum of YD4 from the four main stellar populations: 0-5 (red), 5-10 (blue), 35-70 (pink) and $>75$ Myrs (purple). This comparison allows us to understand why \texttt{PROSPECTOR} preferentially obtains a rejuvenating SFH. The youngest age bin contributes almost all of the nebular emission line flux, while the age bins above 35 Myrs dominate the continuum emission. The 5-10 Myrs bin contributes to the emission line fluxes, however this contribution is significantly smaller than that from the 0-5 Myr bin, even when normalising by the stellar mass in each bin. This decreasing emission line flux with age, and increasing Balmer break strength with age provides an explanation for the rejuvenating solution. Both a young and old stellar population are required to provide the emission lines and Balmer break, respectively. However an intermediate stellar population would both act to drown out the Balmer break, but equally importantly would reduce the EW of the emission lines. In order to reproduce the observed equivalent widths when introducing an intermediate age stellar population, one would need to increase the relative contribution of the youngest age bins. However, this in turn would act to reduce the Balmer break strength, thus requiring a larger old stellar population. Therefore, a careful ratio of the young to old age stellar populations is required with a very small intermediate age stellar population. 

Our \texttt{PROSPECTOR} fitting strongly supports the conclusion we reach using \texttt{CLOUDY} -- that these galaxies are chemically enriched and that a low ionisation parameter is seen. Moreover, they provide an explanation for this evolutionary state -- that these galaxies have previously hosted a large stellar population that is now mature and has hence enriched their ISM. \texttt{PROSPECTOR} also indicates a strong preference for the young stellar population to be a very small fraction of the total stellar mass. This is likely driven by the fact that a relatively low-mass, young stellar population is considerably brighter than older stellar populations of considerably greater stellar mass. This results in an ``outshining'' effect of the old stellar population \citep[see][]{Papovich+01, Pforr+12, Conroy+13, Tacchella+23, Whitler+23}. Further confounding this problem is the dominance of inverse Balmer jumps in the spectra of high-redshift, young galaxies \citep[e.g.][]{Roberts-Borsani+24} that will act to cancel out the Balmer break produced by old stellar populations. 

\begin{figure}
\centering
\includegraphics[width=0.45\textwidth]{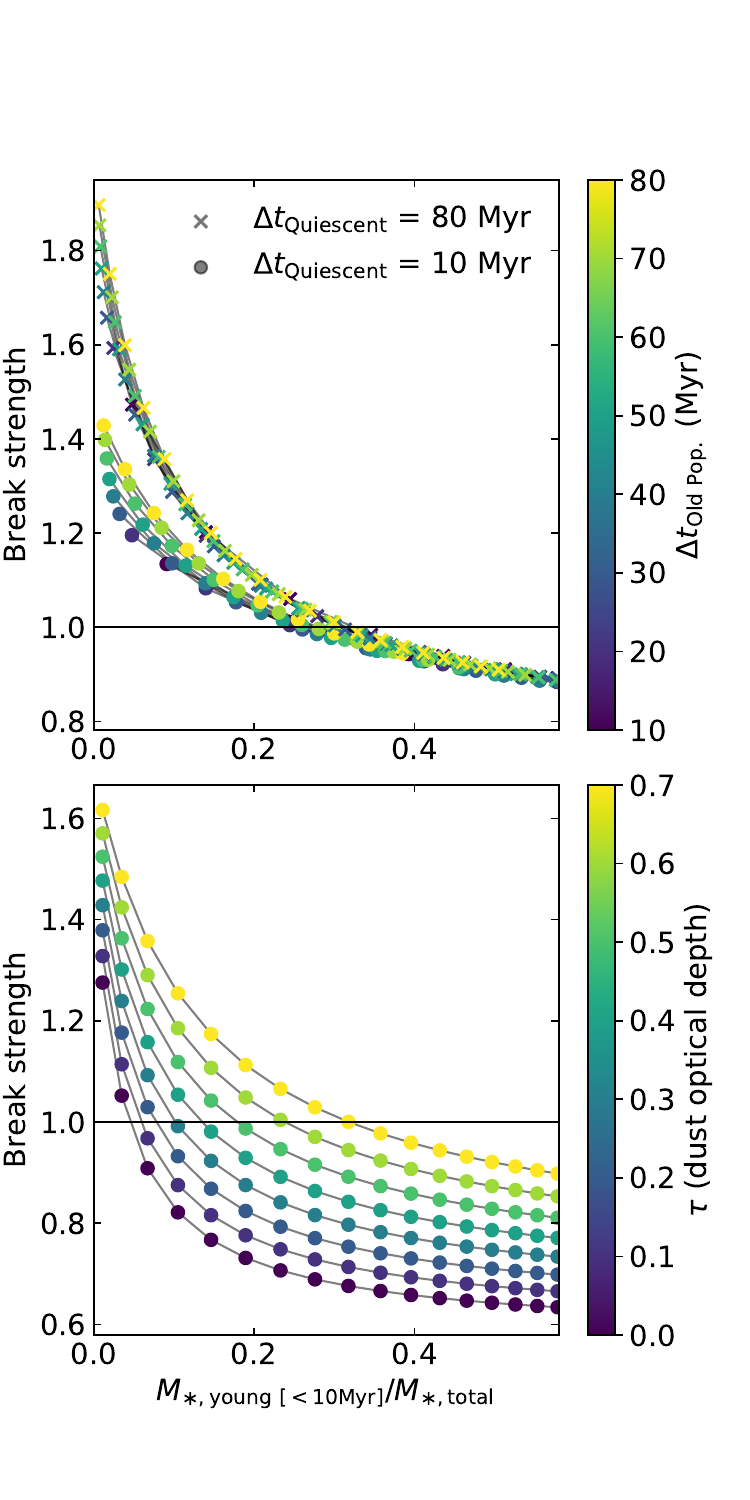}
\caption{Balmer break strength, as defined by \citet{Roberts-Borsani+24}, as a function of the fraction of the total stellar mass produced in the most recent 10 Myr. (Top panel) The effects of changing the time the galaxy spends in a quenched phase (different markers) and the time over which the old stellar population formed (colorbar). The model utilises the galaxy properties of YD4 reported in Section~\ref{sec:SED-fitting}, while varying the SFH. We vary both the time bins, three bins spanning 0--10, 10--(10+$\Delta t_{\rm{Quiescent}}$) and (10+$\Delta t_{\rm{Quiescent}}$)--(10+$\Delta t_{\rm{Quiescent}}$+$\Delta t_{\rm{Old.\ Pop.}}$) Myrs, and the mass within each bin, $M_{\star , \rm{young [< 10Myr]}}$, 0 and $(M_{\star , \rm{total}} - M_{\star , \rm{young [< 10Myr]}})$, respectively. When the young stellar population hits $\sim30\%$ of the total stellar mass, it completely outshines the old stellar population. (Lower panel) The impact of dust attenuation (colorbar) on the observed Balmer break strength. We exploit the properties and SFH of YD4 inferred from our \texttt{PROSPECTOR} SED-fitting while varying the dust V-band optical depth and the ratio of stellar mass formed in the two star formation bursts shown in Figure~\ref{fig:SFHs}. Increasing the amount of dust increases the observed Balmer break. Failing to account for dust will therefore bias the age estimate, however, our \texttt{PROSPECTOR} SED-fitting finds a high value of $\tau \sim 0.7$ and hence we are indeed accounting for significant dust attenuation. Given this, the inferred age of YD4 from \texttt{PROSPECTOR} is likely a conservative estimate. An increasing young stellar population reduces the Balmer break strength at a similar rate regardless of the value of $\tau$.
}
\label{fig:Outshining}
\end{figure}

In this context, we investigate how a young stellar population can wash out a break, by modelling rejuvenating galaxy spectra with the SED-fitting code \texttt{PROSPECTOR} \citep{Prospector}. We create model spectra for a galaxy with the same properties (e.g. gas-phase and stellar metallicity, logU and dust properties) as YD4, as determined by \texttt{PROSPECTOR}, as described above. In order to investigate the "outshining" abilities of young stellar populations, we vary the fraction of the total stellar mass formed within the last 10 Myrs (the duration of the most recent star forming burst). We then enforce a quiescent period with duration $\Delta t_{\rm{Quiescent}}$. This is preceded by the first star formation event, a burst lasting for $\Delta t_{\rm{Old \ Pop.}}$. The varying effects of these can be seen in the upper panel of Figure~\ref{fig:Outshining}. The results of this show that regardless of the time spent in a quiescent stage a young stellar population with a mass fraction of $30\%$ can effectively remove the presence of a Balmer break in a galaxy like YD4. With a shorter period of time spent quiescent, the duration over which the old stellar population formed becomes more important given that the strongest Balmer breaks are produced by stars that are $t \gtrsim 50$ Myrs old. However, as we move towards a lengthy period of time spent in a quiescent state, all of the old stellar population produce large Balmer breaks, regardless of how close to the beginning of the quiescent period that they formed, removing any relation with $\Delta t_{\rm{Old \ Pop.}}$. 

In the lower panel of Figure~\ref{fig:Outshining} we emphasise the impact that varying the diffuse dust V-band optical depth, $\tau$, has on the Balmer break. In order to estimate the impact of varying $\tau$ we create models using the properties and SFH of YD4 that are a result of our \texttt{PROSPECTOR} modelling above. As we decrease $\tau$ from the value observed in YD4 ($\tau \sim 0.7$) we find a decreasing Balmer break strength that is driven by the decreasing attenuation of the UV continuum relative to the optical continuum. However, what becomes apparent from this plot is that even when no dust is present, a Balmer break would be observed in YD4, but only if there is a negligibly small young stellar population. This dust-free Balmer break is "outshone" very efficiently by a young stellar population with a stellar mass of just $\sim 5 \%$. However, given that YD4 has a red UV continuum and ALMA dust detections, we know it hosts a significant mass of dust \citep[$10^{6.5} M_{\odot}$;][]{Hashimoto+23}, thus constraining us to the high optical depth regime. Combining this with the knowledge that YD4 exhibits strong nebular emission lines but also a Balmer break ($B\sim1.3$), evidencing the presence of a young stellar population and a significant old stellar population, we can identify that YD4 can only have a young stellar population with stellar mass fraction of $\lesssim 10\%$. Reducing the quiescent period length, as in the upper panel of Figure~\ref{fig:Outshining}, can only act to decrease the implied stellar mass of the young stellar population, in order to retain the break strength. However, the presence of strong nebular emission lines helps \texttt{PROSPECTOR} to further constrain the fraction of total stellar mass from the young stellar population to be $\sim 5\%$, as discussed further above. 

The implication therefore is that YD4 and our stack host both young and old stellar populations, but are in a very short-lived phase where these are both identifiable. With a larger young stellar population ($\sim 30\%$), the Balmer break seen in YD4 would be entirely lost, while with a slightly more mature young stellar population the emission lines would not be seen. 

\subsection{SPHINX}
\label{sec:SPHINX}

\begin{figure}
\centering
\includegraphics[width=0.5\textwidth]{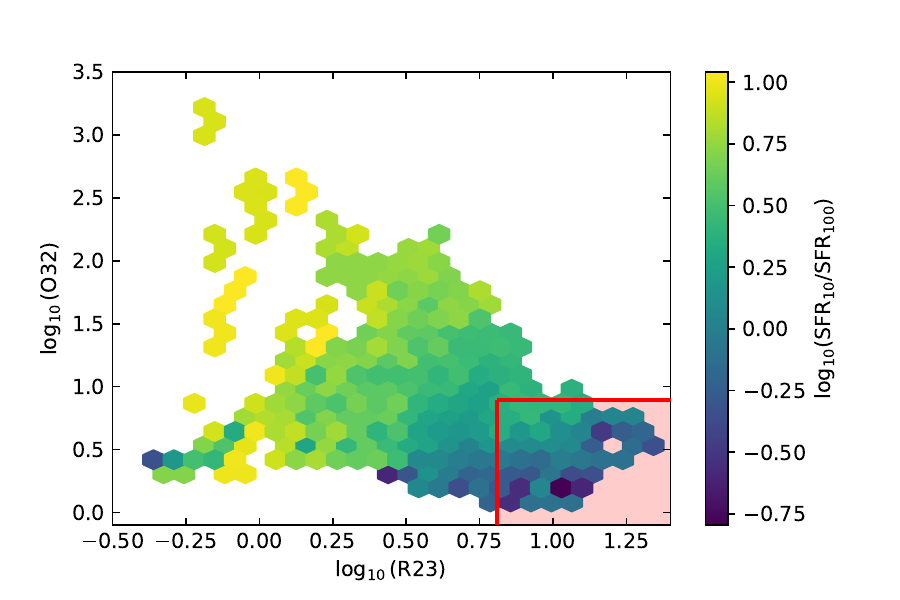}
\caption{The O32 ratio against R23 ratio of mock observed galaxies, color coded by the average ratio of the recent SFR over 10 Myr and the SFR over 100 Myr in each hexagonal bin. These mock observed galaxies are taken from the SPHINX$^{20}$ simulation, which includes the effects of dust attenuation \citep[see ][]{Rosdahl+18,Rosdahl+22, Katz+23}. The ratios are shown for all sightlines of all galaxies at $z\geq6$. There is a clear correlation between a decline in the ongoing SFR (i.e. a low $\rm{SFR}_{10}$/$\rm{SFR}_{100}$) and the emission line ratios we target, facilitating the observation of the Balmer break. The red shaded area (background) and red box (foreground) indicates the same region as in Figure~\ref{fig:exc_vs_ion} but before accounting for dust corrections. A trend towards a decrease in recent star formation is seen in SPHINX$^{20}$ galaxies towards the location of our sample of low O32 and high R23 sample of galaxies, as expected from galaxies that have recently experienced a mini-quenched period.}
\label{fig:SPHINX}
\end{figure}

\begin{figure*}
\centering
\includegraphics[width=1\textwidth]{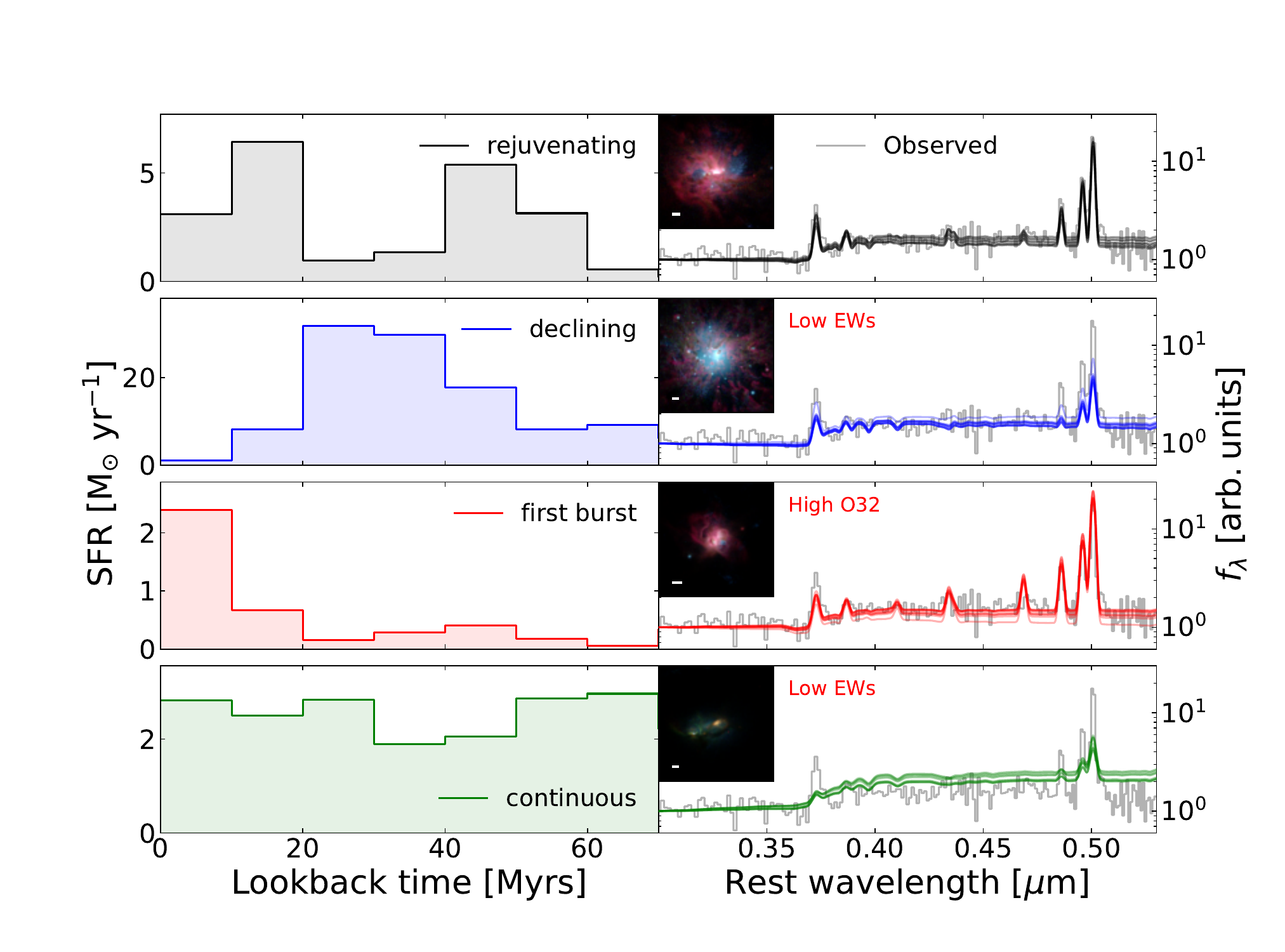}
\caption{SPHINX simulated galaxies showing the effects of different SFHs on the observed spectrum: rejuvenating (black), declining (blue), first burst (red) and continuous (green). Left panels show the SFH of each galaxy, where the type of SFH is indicated in the legend. Each SFH has been re-binned to show the average SFR within a 10~Myr bin, variations below this resolution will not significantly effect the observed spectrum and hence are not shown for clarity. Right panels show the corresponding spectra of the galaxy along the 10 lines-of-sight available. The spectrum of YD4 is shown in grey for comparison. All spectra are normalised by their UV flux at 0.15 $\mu$m. The inset panel shows the RGB image along one line-of-sight, using mock F150W, F277W and F444W images, of the SPHINX galaxies, where the white bar indicates a length of 1~pkpc. The only SFH that recreates both the emission line ratios and equivalent widths seen in YD4 are those that host a recent rejuvenation event. The rejuvenating galaxy and the declining galaxy both show large spatial extents relative to the first burst and continuous SFH galaxies.}
\label{fig:SPHINX_SFH}
\end{figure*}

Given the combination of an evolved galaxy hosting a young stellar population, that is growing up in a chemically enriched and disrupted ISM, we employ a more complex comparison to the SPHINX cosmological radiation-hydrodynamic simulations \citep{Rosdahl+18,Rosdahl+22}. We make use of the SPHINX$^{20}$ public data release \citep{Katz+23}. The detailed modelling of the multi-phase ISM in SPHINX is ideally suited to understanding the evolution of emission lines originating from evolved galaxies. In this context, we first consider the ratio between the recent SFR over 10 Myrs and the SFR over 100 Myrs. If our conclusion regarding the required relatively low mass of the young stellar population in these rejuvenating galaxies is accurate we would expect a strong dependence on SFR$_{10}$/SFR$_{100}$ to be seen in the excitation-ionisation plot of SPHINX$^{20}$ $z\gtrsim 6$ galaxies. We show the excitation-ionisation plot, using dust-attenuated emission line ratios, in Figure~\ref{fig:SPHINX} and demonstrate there is a strong correlation between a decrease in SFR and the emission line ratios we target. 

We also wish to assess whether the SPHINX galaxies that fall within the same selection criteria used in Section~\ref{sec:stack} can replicate the observed spectrum of YD4. In Figure~\ref{fig:SPHINX_SFH} we present four different regimes of SFHs seen in the SHPINX simulations. The upper panel shows the SFH and spectrum of a rejuvenating galaxy. The SPHINX snapshot catches the galaxy on the downturn of its rejuvenation event, but just recently enough that the emission lines are still visible. We note that catching this galaxy at a slightly earlier epoch, during its burst, would help to boost the EW of the emission lines that are currently slightly lower, but still within the errorbars of the JWST spectrum, than those seen in YD4. This SFH is able to replicate the spectrum of YD4 within its uncertainties, both in terms of emission line ratios and EWs, as well as the Balmer break strength. Conversely, the declining SFH produces a spectrum that has emission lines that have much lower EWs than those seen in YD4, but the emission line ratios are well replicated, suggesting that the emission line ratios are indeed driven by the effects on the ISM of an old stellar population. The ``first burst'' galaxy shows a SFH that is dominated by a single burst of star formation that is ongoing. While the presence of star formation over $\sim 100-200$~Myr before observation accounts for the observed Balmer break, the emission lines from the dominant burst fail to replicate the emission line ratios of YD4, with too little \OII\ emission and too much \OIII\ emission. Finally, the continuous SFH produces a significant Balmer break but fails to reproduce the emission lines. The only galaxy that reproduces the emission line ratios, equivalent widths and Balmer break of YD4 is the galaxy that shows a recent rejuvenation event following a period of quiescence that lasted $\sim 20$ Myr, in strong agreement with the SFH of YD4 that is produced with \texttt{PROSPECTOR} SED-fitting.

However, what is clear is that the young stellar population in the rejuvenating galaxy we took from the SPHINX$^{20}$ simulation has a considerably larger fraction of the stellar mass built up in the $0-70$ Myr history, shown in Figure~\ref{fig:Outshining}, than we derived for YD4. This is driven by the more complex SFH that is modelled in the SPHINX$^{20}$ simulation, which shows an extended SFH over some hundreds of Myrs that helps to build up stellar mass and the Balmer break.

\begin{figure}
\centering
\includegraphics[width=0.5\textwidth]{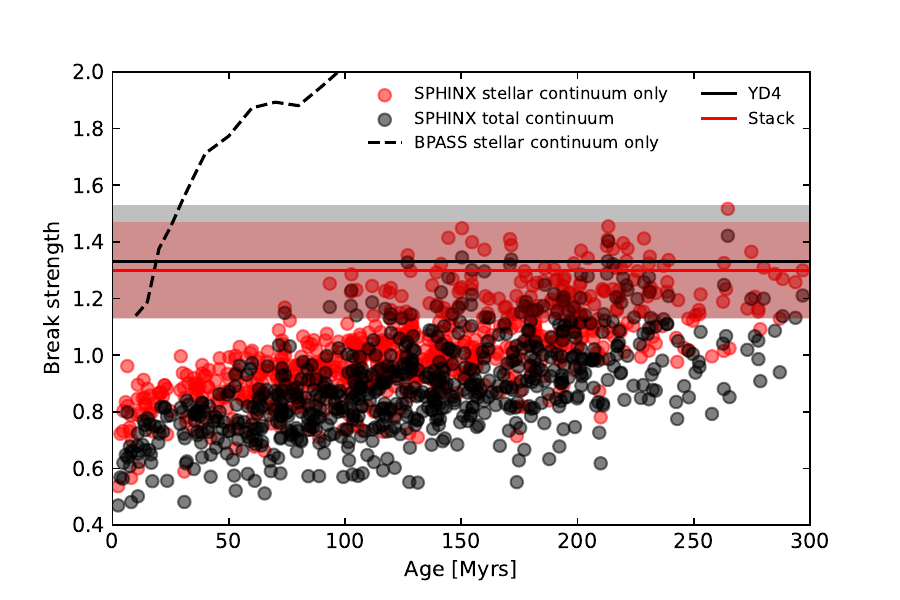}
\caption{The Balmer break strength as a function of stellar-mass-weighted age in the SPHINX$^{20}$ public data release \citep{Rosdahl+18,Rosdahl+22, Katz+23} at $z\geq 5$. The break strengths seen in the stellar continua (red) and total (stellar and nebular) continua (black) are shown. The Balmer break strength seen in the stellar continuum of the BPASS models with a metallicity of log$_{10}Z/Z_{\odot} = -0.5$, as a function of the age of the model, is shown by the black dashed line. Finally, the black and red solid lines and shaded regions indicate the break strength and its associated uncertainty for YD4 and the stacked spectrum, respectively.}
\label{fig:SPHINX_breaks}
\end{figure}

Therefore, we consider the Balmer break strength, using the definition discussed in Section~\ref{sec:YD4}, seen in SPHINX$^{20}$ galaxies as a function of mass-weighted age in Figure~\ref{fig:SPHINX_breaks}. The break strength observed in the stellar continuum is, unsurprisingly, higher than that observed in the total galaxy spectrum. This is driven by the presence of inverse Balmer breaks in the nebular component of the total spectrum reducing the total break strength. This effect can also be seen in the break strengths of the BPASS SEDs. These breaks are considerably stronger than those seen in the SPHINX$^{20}$ galaxies due to our purposeful neglect of nebular emission in our BPASS break strength measurement. These models use a single stellar population of a given age, however the SPHINX$^{20}$ galaxies are accumulations of various stellar populations of different ages. Young stellar populations show no evidence of Balmer breaks in their stellar continua and moreover dominate the total galaxy spectrum. This results in the significant reduction in the break strength of galaxies composed of multiple stellar populations. Thus, when we compare to the more realistic SPHINX$^{20}$ galaxies, the break strengths seen in YD4 and the stack suggest that these galaxies host stellar populations with ages between $75-300$ Myrs. The mass weighted ages of YD4 and the stack implied from their SFHs are 80 Myrs and 70 Myrs, respectively. While modelling differences may drive this inconsistency between the ages implied from \texttt{PROSPECTOR} and SPHINX$^{20}$, the ease for a relatively young stellar population to wash out a Balmer break (shown in Figure~\ref{fig:SPHINX_breaks} and discussed further in Appendix~\ref{sec:Looser+23}), the effect known as outshining \citep[e.g.][]{Papovich+01, Pforr+12, Conroy+13, Tacchella+23, Whitler+23}, is likely to be playing a crucial role here. This degeneracy means that we are potentially missing significant amounts of stellar mass when we fit these apparently young, bursty galaxies with SED-fitting codes. This is again underlined by the SPHINX$^{20}$ prediction that YD4 and the stack host older stellar populations than is predicted by \texttt{PROSPECTOR} -- only the most recent major burst can be identified based on the presence or lack of a Balmer break. This implies that YD4 has a formation redshift of at least $z\sim 9$ and potentially as early as  $z\sim 12.5$, underlining the challenges that such bursty star formation histories provide when trying to diagnose the age and stellar mass of high-redshift galaxies. 

\section{Discussion and conclusion}
\label{sec:Conclusion}
In this paper we have presented JWST NIRSpec PRISM spectra that show both a Balmer break and emission lines. These results are significant in the context of recent detections of (mini-)quenched galaxies at $z\gtrsim5$ by JWST \citep{Looser+23,Looser+23b,Dressler+23,Dressler+24,Strait+23,Trussler+24}. These galaxies have been observed to host Balmer breaks without the presence of emission lines. We have established that by selecting galaxies in the high metallicity and low ionisation region of the R23-O32 plane, the median galaxy at $z>5.5$ hosts a Balmer break, in strong contrast to galaxies with higher O32 ratio, as well as the general star-forming galaxy population. We thus conclude that these abnormal emission line ratios are indeed linked to the presence of an old stellar population. \texttt{CLOUDY} modelling helps us to conclude that this appears to be driven by the chemically enriched and disrupted ISM that is produced by these older stellar populations. These results strongly imply the presence of both an old stellar population ($\sim 100$ Myrs) producing the Balmer break and a young stellar population ($\lesssim 10$ Myrs) driving the observed strong nebular emission lines. 

These results naturally lead us to query what is the shape of the SFH producing these observed spectral properties. We employ a non-parametric SFH modelling approach, exploiting the SED-fitting code \texttt{PROSPECTOR}. The results of this strongly favour the conclusion that both YD4 and our stack are best described by a rejuvenating SFH. Our modelling of the Balmer break strength strongly constrains the young stellar population to be less than a tenth of the total stellar mass, while the strength of the observed emission lines constrains this fraction to close to $5\%$. Following this, we exploit the SPHINX$^{20}$ simulations to further constrain the range of SFHs that can replicate the spectral features seen from YD4 and our stack. SPHINX$^{20}$ galaxies show that a rejuvenating SFH is able to match the break strength, emission line ratios and EWs. Alternative SFHs can replicate some aspects of the observed spectrum of YD4 and our stack, but fail to match them all simultaneously.

While the declining SFHs in SPHINX$^{20}$ fail to match the observed spectrum of YD4, we do believe that a carefully tuned gradual decline in star formation could feasibly result in a young stellar population that produces nebular emission lines alongside older stars which produce a Balmer break. Moreover, a galaxy with this star-formation history would host an enriched ISM, which is consistent with the emission line ratios. However, the high observed EW of the emission lines and the strength of the Balmer break place strong constraints on the required young and old stellar populations. As discussed in Section~\ref{sec:SED-fitting}, significant intermediate star formation (10--50\,Myr) would act to reduce the EW of emission lines and start to wash out a large Balmer break, in conflict with the observed data. Additionally, the strong preference of our SED fitting for little intermediate star formation may be informed by other constraints, such as the UV and optical slopes. We also note that this more gradual decline scenario is at odds with the rapid quenching timescales which have been inferred for other galaxies in the early Universe \citep{Looser+23b, Gelli+23, Dome+23, McClymont+24b}. 

All of the results of our modelling and analysis of the spectra of YD4, our stack and SPHINX$^{20}$ galaxies strongly support the premise that we are observing rejuvenation in a galaxy with a chemically enriched, disrupted ISM. We find that a significant population of galaxies exist that host both young and old stellar populations evidencing their complex SFHs. These spectral properties are best explained by galaxies hosting multiple bursts of star formation. Moreover, bursty star formation may need to be invoked to understand the overabundance of UV bright galaxies in the early Universe \citep[e.g.][]{Kravtsov+24, Sun+23, Mason+23} and may in part be a result of the rapid build up of stellar mass through galaxy mergers at extremely high redshifts, as seen in \cite{Witten+24}. 

It is notable that the SFH of YD4 and our stacked spectrum, inferred from \texttt{PROSPECTOR} SED-fitting, implies that both must have experienced a quiescent period of $\sim 25$ Myr. Moreover, the rejuvenating galaxy taken from the SPHINX$^{20}$ simulations also shows a reduction in star formation over a 20 Myr timescale. This conclusion about the short timescales of quenching events has been suggested by previous works \citep[e.g.][]{Dome+23, Looser+23b, McClymont+24b}. The timescales between bursts is regulated by the mode of quenching. New star-formation events are as a result of either the in-fall of gas onto the galaxy, or the return of once feedback-affected gas to a cool and dense state within the galaxy, such that new stars can be formed. As a result, a wide range of quenching mechanisms exist: a lack of gas accretion from the CGM, a lack of gas cooling within the ISM, mechanisms stopping the cold gas from forming stars, the rapid consumption of the available cold gas within a galaxy or the removal of gas from the ISM \citep[see][]{Man&Belli}. The short mini-quenched timescales inferred from our modelling of the spectra of YD4 and our stack suggests that the rapid consumption of all gas or strong AGN feedback scenarios are unlikely. This is further supported by the lack of evidence of AGN activity. Moreover, if mergers were driving the removal or rapid consumption of cold gas within these mini-quenched galaxies, and in turn driving the rejuvenation of star formation, from our quenching timescales we would expect galaxy mergers to be occurring at a frequency of one every few tens of Myr while observational evidence suggests the frequency is an order of magnitude less often (Pusk\'{a}s et al. in prep.). Instead, disruptive feedback driven by star formation or weak AGN feedback is more likely to quench galaxies on these short timescales. Through comparison with multiple simulations \cite{Dome+23} find that $z=4-7$ mini-quenched galaxies are typically in a quenched state for 20-40 Myrs. They identify that this is consistent with the free-fall timescale of the inner halo, which further supports the conclusion that stellar feedback or weak AGN feedback is crucial in producing these bursty SFHs. These processes expel gas from the ISM leaving the galaxy in a mini-quenched phase for a few 10's of Myrs, after the free-fall time of the inner halo, this gas reaccretes and rejuvenates the galaxy. 

This growing evidence that galaxies in the early Universe are composed of stellar populations produced by multiple bursts of star formation underlines the importance of accounting for this in our modelling of their SFHs. However, our modelling also shows the fragility of the Balmer break strength as a tracer of old stellar populations, due to the ease at which it can be drowned out by young stars. While the presence of a Balmer break can be taken as evidence of an old stellar population, its absence cannot be used to imply a lack of one. The effects of stochastic SFHs will act to drown out the optical continuum of old stellar populations and hence will bias us towards underestimating the age and stellar mass of these galaxies in the very early Universe. 

\section*{Acknowledgements}
The authors would like to thank Francesco Belfiore, Roser Pell\'o and Ilias Goovaerts for enlightening conversations. 

CW thanks the Science and Technology Facilities Council (STFC) for a PhD studentship, funded by UKRI grant 2602262. WM thanks the Science and Technology Facilities Council (STFC) Center for Doctoral Training (CDT) in Data intensive Science at the University of Cambridge (STFC grant number 2742968) for a PhD studentship. ST acknowledges support by the Royal Society Research Grant G125142. DS acknowledges support by the Science and Technology Facilities Council (STFC). CS, JW, RM, FDE, XJ and TJL acknowledge support by the Science and Technology Facilities Council (STFC) and by the ERC through Advanced Grant number 695671 ‘QUENCH’, and by the UKRI Frontier Research grant RISEandFALL. RM also acknowledges funding from a research professorship from the Royal Society. TJL further acknowledges support by the STFC Center for Doctoral Training in Data Intensive Science.

\section*{Data Availability}
All JWST and HST data products are available via the Mikulski Archive for Space Telescopes (https://mast.stsci.edu). The spectra of our stacks are available upon request. 
\bibliographystyle{mnras}
\bibliography{example} 

\appendix
\section{Analysis of mini-quenched source}
\label{sec:Looser+23}

\begin{figure}
\centering
\includegraphics[width=0.5\textwidth]{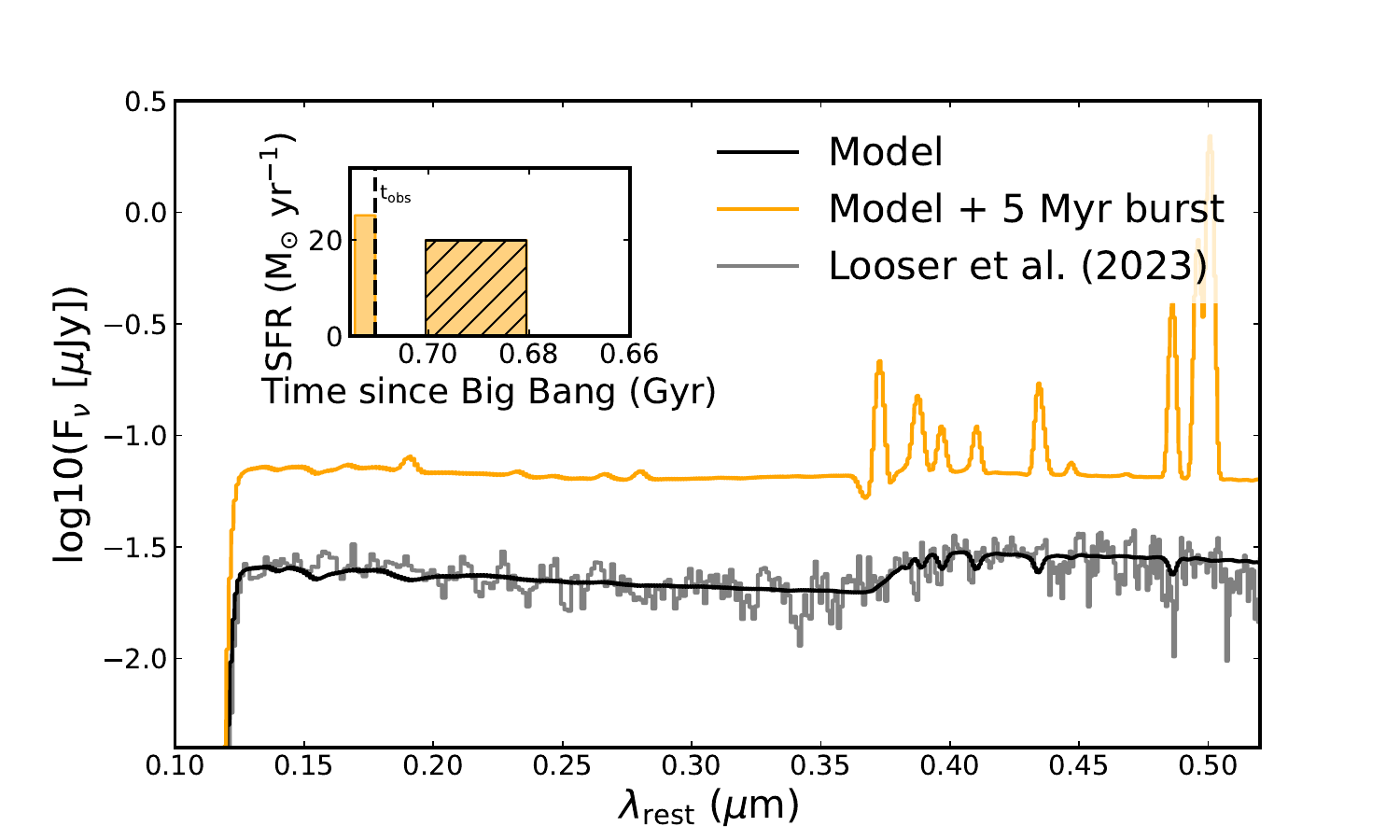}
\caption{The NIRSpec/JWST observation of the suspected (mini-) quenched galaxy from \citet{Looser+23} is shown in grey. No emission lines are detected and a clear Balmer break is seen. The black curve shows the \texttt{BAGPIPES} model spectrum of this galaxy, while the yellow curve shows the same model galaxy but with a burst ($\Delta t = 5$ Myrs) appended to its SFH as shown in the inset panel. In this case, the Balmer break is completely drowned out by the young stellar population.}
\label{fig:L23}
\end{figure}

In order to assess whether the spectrum of the quiescent galaxy at $z\sim7$ in \cite{Looser+23} can also be drowned out by a similar size young stellar population ($\sim 30\%$, as discussed further in Section~\ref{sec:SED-fitting}) we perform SED-modelling exploiting the SED-fitting code \texttt{BAGPIPES} \citep{BAGPIPES}. The SFH presented in \cite{Looser+23} shows a burst of star formation occurring over a period of $\sim 30$ Myrs followed by a period of quiescence lasting $\sim 10$ Myrs. We take the spectrum, properties and SFH of the \texttt{BAGPIPES} fit presented in \cite{Looser+23}, seen in Figure~\ref{fig:L23}, and apply a burst of star formation for 5 Myrs post-observation. Within this period the galaxy returns to its pre-quenching SFR and increases its stellar mass by $\sim 30 \%$. The Balmer break, originally present in the spectrum of \cite{Looser+23}, is completely drowned out by this relatively small, young stellar population in strong agreement with our analysis of Balmer break strengths with \texttt{PROSPECTOR}. 

\bsp	
\label{lastpage}
\end{document}